\documentclass[aps,pra,twocolumn,superscriptaddress,floatfix]{revtex4}
\usepackage{epsfig,color,graphicx}
\usepackage{amsmath,amssymb,amsfonts,amsthm}
\begin{document}

\title{Expansion velocity of a one-dimensional, two-component Fermi gas during the sudden expansion in the ballistic regime}

\author{S. Langer}
\affiliation{Department of Physics and Arnold Sommerfeld Center for Theoretical Physics, Ludwig-Maximilians-Universit\"at M\"unchen, D-80333 M\"unchen, Germany}
\author{M.J.A. Schuetz}
\affiliation{Department of Physics and Arnold Sommerfeld Center for Theoretical Physics, Ludwig-Maximilians-Universit\"at M\"unchen, D-80333 M\"unchen, Germany}
\affiliation{Max-Planck-Institut f\"ur Quantenoptik, Hans-Kopfermann-Strasse 1, 85748 Garching, Germany}
\author{I.P. McCulloch}
\affiliation{Centre for Engineered Quantum Systems, School of Mathematics and Physics, The University of Queensland, St Lucia, QLD 4072, Australia}
\author{U. Schollw\"ock}
\affiliation{Department of Physics and Arnold Sommerfeld Center for Theoretical Physics, Ludwig-Maximilians-Universit\"at M\"unchen, D-80333 M\"unchen, Germany}
\affiliation{Kavli Institute for Theoretical Physics, Kohn Hall, University of California, Santa Barbara, California 93106, USA}
\author{F. Heidrich-Meisner} 
\affiliation{Department of Physics and Arnold Sommerfeld Center for Theoretical Physics, Ludwig-Maximilians-Universit\"at M\"unchen, D-80333 M\"unchen, Germany}
\affiliation{Kavli Institute for Theoretical Physics, Kohn Hall, University of California, Santa Barbara, California 93106, USA}
\date{\today}

%--------------------------------------------------------------------------
\begin{abstract}
We show that in the sudden expansion of a spin-balanced, two-component Fermi gas into an empty optical lattice 
induced by releasing particles from a trap, over a wide parameter regime, the radius $R_n$ of the particle cloud 
grows linearly in time. 
This allow us to define the  expansion velocity $V_{\mathrm{ex}}$ from $R_n=V_{\mathrm{ex}}t$.
The goal of this work is to clarify the dependence of the expansion velocity on the initial conditions which we establish
from time-dependent density matrix renormalization group simulations, both for a box trap and a harmonic trap. As a prominent result,  the presence  of a Mott-insulating 
region leaves clear fingerprints in the expansion velocity. Our predictions can be verified in experiments with ultra-cold atoms. 
\end{abstract}
%--------------------------------------------------------------------------

\maketitle
%**************************************************************************
% Introduction
%**************************************************************************

\section{Introduction}
Research into the non-equilibrium properties of strongly correlated many-body systems
has  emerged into a dynamic and active field, driven by the possibility to address questions such as
 thermalization \cite{rigol08a,polkovnikov11}, the properties of steady states, or state engineering in ultra-cold atomic gases \cite{bloch08}.
While substantial theoretical attention has been devoted to quantum quenches  in homogeneous systems
\cite{polkovnikov11}, more recently, 
set-ups that give rise to finite particle or spin currents have been studied as well, both from the
theoretical side \cite{rigol04,rigol05b,rosch08,mandt11,hm08a,hm09a,muth12,hen10,jreissaty11,lundh11,delcampo06,delcampo08} and in experiments (see, {\it e.g.}, Refs.~\onlinecite{anker05,kinoshita06,schneider11,medley11,sommer11b,joseph11}).
Using these approaches  allows one to investigate transport properties of strongly correlated many-body systems - in and out-of-equilibrium - in cold atomic gases
that are  of great interest in condensed matter theory. 

Our work is motivated by  the experiment by Schneider {\it et al.}~\cite{schneider11} who have studied the expansion
of a two-component Fermi gas in an optical lattice in two and three dimensions (described by the Fermi-Hubbard model \cite{schneider08,joerdens08}), starting from 
 an almost perfect band insulator. 
The qualitative interpretation of their results is that,  besides a ballistically propagating
halo of particles, at finite interaction strengths a core of diffusively expanding particles exists \cite{schneider11}. 
In the case of one-dimensional (1D) bulk systems relevant for condensed matter problems and on the level of linear response theory,  ballistic dynamics of interacting particles can be traced back to the existence of non-trivial conservation laws \cite{zotos97}.
For instance, the fact that the energy current is conserved for the 1D Heisenberg model renders its spin transport
ballistic away from zero total magnetization \cite{zotos97,transport1D,rosch00}, whereas at zero magnetization 
there exists a quasi-local quantity  \cite{prosen11}, which  is conserved only for the infinite system, 
that gives rise to ballistic dynamics.
While for the 1D Hubbard model, the understanding of its transport properties is by far less complete than for 
the Heisenberg chain, one might be tempted to expect similar quantities to play a role for the latter model as well \cite{zotos97}.  

A qualitative difference between the sudden expansion in an optical lattice compared to steady-state transport measurements in condensed matter systems
is that, in the latter case, the background density determines transport coefficients, whereas in the former case, the density itself becomes time-dependent \cite{schneider11}
and {\it all} particles participate in the dynamics. As a consequence, in diffusive regimes, the dependence of the diffusion coefficient on  density needs to be accounted for.
In the ballistic case, as we shall see, the expansion velocity always depends on all momenta that are occupied in the initial state and not on just those close to the
Fermi wave-vector.
Therefore, a parameter regime complementary to condensed matter systems can be accessed with cold atoms.

Theoretical results for the expansion of interacting  bosons or fermions  in optical lattices are mostly available for the 1D case,
for which exact numerical methods give  access to at least the short time dynamics via the adaptive time-dependent density matrix renormalization group (tDMRG) method 
\cite{daley04,white04,schollwoeck05,schollwoeck11} or exact diagonalization (ED) \cite{rigol04,rigol05b}.
The richness of the non-equilibrium physics encountered in the expansion manifests itself in the observation
of the dynamical emergence of coherence \cite{rigol04,rodriguez06,hm08a,hen10}, which, for bosons, leads to the phenomenon of  dynamical quasi-condensation \cite{rigol04,rodriguez06,hen10} and the
intriguing phenomenon of the fermionization of the momentum distribution function (MDF)  \cite{rigol05b,minguzzi05,delcampo08,gritsev10}.
In the case of a two-component Fermi gas,  the short-time dynamics of the MDF and correlation functions \cite{hm08a},
the emergence of  metastable states \cite{hm09a,muth12} and the time-evolution of density profiles for specific initial conditions
have  been investigated  \cite{hm08a,hm09a,karlsson11,kajala11,kajala11a}.

In the present work we study the 1D Hubbard model and
we concentrate on the sudden expansion starting from initial states that are  Mott insulators (MI), {\it i.e.}, that have an integer filling of $n_{\mathrm{init}}=1$, 
Tomonaga-Luttinger (TL) liquids ($n_{\mathrm{init}}<1$), or systems in a harmonic trap. 
In the latter case,  depending on filling and interaction strength, several phases may coexist in  separate shells \cite{rigol04b}. 
We analyze the dependence of the expanding cloud's radius $R_{n}(t)$ on time $t$ and search
for conditions to obtain ballistic dynamics, for which $R_n(t)=V_{\mathrm{ex}}\, t$ is a necessary criterion.
In that case, the expansion velocity $V_{\mathrm{ex}}$ is a well-defined quantity, and, as a key result of our work, we clarify 
its dependence on the initial conditions.

\begin{figure}[t]
\includegraphics[width=0.45\textwidth]{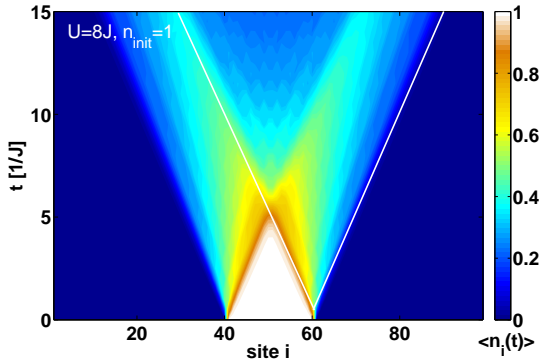}
\caption{(Color online)
Box trap: Typical contour plot of the density $\langle n_i(t)\rangle $ during the
expansion from a MI   ($U=8J$, $n_{\mathrm{init}}=1$, $L_{\mathrm{init}}=20$). The slanted lines indicate the speed $2J $ at which the
MI melts.}\label{fig:contour}
\end{figure}

Our main results are: (i) 
In the regime of low densities, {\it i.e.},  $n_{\mathrm{init}} \leq 1$, 
we  observe a linear growth 
of the cloud's radius with time, allowing us to define  $V_{\mathrm{ex}}$. 
(ii) In general, the expansion speed $V_{\mathrm{ex}}$ 
depends in a non-monotonic way on the initial density. In the case of the expansion from a MI, 
  $V_{\mathrm{ex}}=\sqrt{2}J$, independently of $U$. 
(iii) Our findings are robust against the presence of a harmonic trap in the initial state.

Note that, in a generic system, one expects ballistic dynamics in the {\it long} time limit, where
the gas becomes so dilute that interactions cease to matter. Here we show that
ballistic dynamics sets in {\it immediately} after the gas is released from the
trap when the density is actually still comparable to the initial density.

The structure of the paper is the following: In Section~\ref{sec:model}, we introduce the model and define the radius of the cloud. Section~\ref{sec:box} discusses the expansion from a box trap,{\it i.e.}, starting from a homogeneous density. We first show that the dynamics is ballistic by analyzing the radius and the particle currents and second, we present a detailed analysis of the expansion velocity as a function of density and interaction strength. In Sec.~\ref{sec:trap} we test our findings against the inhomogeneity introduced by a harmonic trap. We summarize our findings in Sec.~\ref{sec:summary}.
In Appendix A, we discuss the diffusion equation in one dimension. Appendix B contains a finite-size analysis of the expansion velocity for various cases.
 
\section{Model and setup} 
\label{sec:model}
Our study is carried out for the 1D Hubbard model:
\begin{equation}
H_0 = -J\sum_{i=1, \sigma=\uparrow\downarrow}^{L-1} ( c_{i+1,\sigma}^{\dagger}  c_{i,\sigma}^{ }
+ \mathrm{h.c.} ) + U \sum_{i=1}^{L}
n_{i,\uparrow} n_{i,\downarrow} \,. \label{eq:ham}
\end{equation}
$c_{i\sigma}^{\dagger}$ is a fermionic creation operator with spin $\sigma=\uparrow,\downarrow$ acting on site $i$, $n_{i\sigma}=c_{i\sigma}^{\dagger}c_{i\sigma}$, $n_i=\sum_{\sigma} n_{i\sigma}$, $U$ is the onsite repulsion, and $J$, is the hopping matrix element.  Open boundary
conditions are imposed, $L\gtrsim 100$. is the number of lattice sites, and $N$ the number of particles. 
We set $\hbar$ and the lattice spacing to unity and thus measure time, velocity and particle current in the appropriate units in terms of the hopping matrix element.

We prepare initial states as the ground state of 
$
H=H_0+H_{\mathrm{conf}}
$ \cite{hm08a}.
We consider two cases: First, the expansion from a box trap 
({\it i.e.}, $\langle n_i\rangle \not=0$ for $i_1<i\leq i_2$; $(i_2-i_1)=L_{\mathrm{init}}$, $n_{\mathrm{init}}=N/L_{\mathrm{init}} $) enforced by using $H_{\mathrm{conf}}=\sum_i \epsilon_i n_i$ 
with a large $\epsilon_i \gtrsim U$ for $i\leq i_1;  i_2<i$ and zero otherwise). 
The second example is the expansion from a harmonic trap, for which $H_{\mathrm{conf}}=V\sum_i (i-i_0)^2 n_i$.
We turn off $H_{\mathrm{conf}}$ at $t=0$. In our tDMRG runs, we use  
a Krylov-space 
based method \cite{park86,hochbruck97}, with time steps of $\delta t\,J=0.25$ and we enforce a discarded weight of  $10^{-4}$ or smaller.   

The main quantity of interest is  the radius of the particle cloud that we define via 
\begin{equation}
R_n = \sqrt{\frac{1}{N}\sum_{i=1}^L   \langle n_i\rangle  (i-i_0)^2-R_n^2(t=0)}\,.
\label{eq:rn}
\end{equation}
For the expansion from a box, $i_0=L/2+0.5$.

\begin{figure}[t]
\includegraphics[width=0.45\textwidth]{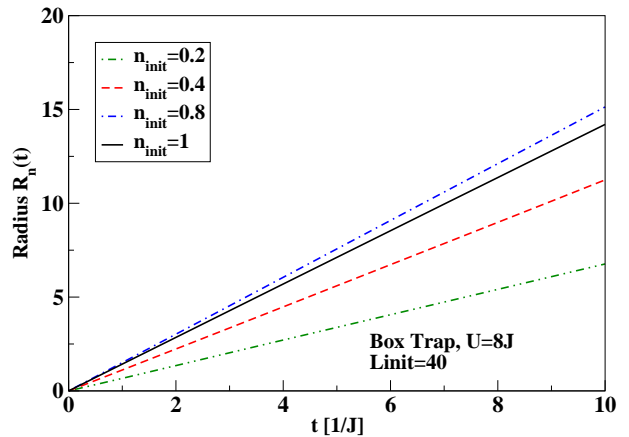}
\caption{(Color online)
Box trap: Radius $R_n(t)$ for initial densities
$n_{\mathrm{init}}=0.2,0.4,0.8,1$ at $U=8J$ and $L_{\mathrm{init}}=40$ (corresponding to $N=8,16,32,40$).}
\label{fig:radius}
\end{figure}

\section{Expansion from a box trap}
\label{sec:box}
We first discuss this idealized case  to avoid 
the complication of dealing with particles originating from different shells, as would be the case
with a harmonic trap (note, though, that box-like traps can also be generated in experiments \cite{ashkin04,meyrath05}). A typical example for the time-evolution of the density $\langle n_i(t)\rangle$
is shown in Fig.~\ref{fig:contour} for the expansion from a MI with $U=8J$. The MI melts on a time scale
of $t_{\mathrm{melt}}\lesssim L_{\mathrm{init}}/(2J)$, where $2J$ is the largest possible velocity in the 
empty lattice, since the single-particle dispersion is $\epsilon_k=-2J\cos(k)$ \cite{hm08a}. For $t>t_{\mathrm{melt}}$, 
two particle clouds form that propagate into opposite directions,
visible as two intense jets (compare Refs.~\cite{polini07,rigol04,rigol05b,langer09,langer11,kajala11}).

In Fig.~\ref{fig:radius}, we display the radius $R_n(t)$ at $U=8J$ for various initial densities at $U=8J$. Clearly,
for $n_{\mathrm{init}}\leq 1$, $R_n(t)=V_{\mathrm{ex}}t $. 
We stress that $R_n(t)\sim t$
sets in immediately after the gas is released from the trap.
This includes, in particular, the expansion from a MI at any $U$, while for $n_{\mathrm{init}}> 1$,
the radius deviates from $R_n(t) \sim t$ \cite{hm09a}. 
Based on the observation of $R_n(t) \sim t$ on short and intermediate times, when local densities are still large,
together with the fact that {\it interacting} particles 
behave similar to non-interacting ones (which, in the absence of disorder, expand with $R_n \sim t$),
we classify the dynamics as ballistic.
 
\begin{figure}[t ]
\includegraphics[width=0.45\textwidth]{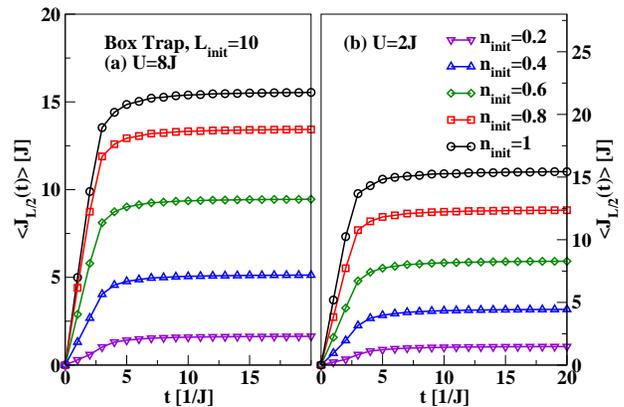}
\caption{(Color online) Box trap: Total particle current in each half of the system as a function of time, i.e.,  $J_{L/2}(t):=\sum_{i>L/2} j_i$, for (a) $U=8J$, (b) $U=2J$ and $n_{\mathrm{init}}=0.2,0.4,0.6,0.8,1$. We observe that after some transient dynamics, $\langle J_{L/2}(t)\rangle=\mathrm{const}$, supporting the observation of ballistic transport. }
\label{fig:currents}
\end{figure}

In our situation, the notion of ballistic dynamics is strongly corroborated by analyzing
the time dependence of the total particle current in each half of the system, $J_{L/2} =\sum_{i>L/2} j_i$ [$j_i=-iJ \sum_{\sigma}(c_{i+1\sigma}^{\dagger}c_{i\sigma}-\mbox{h.c.}) $], which is shown for $U/J=2,8$ in Fig.~\ref{fig:currents}. 
After the two jets in Fig.~\ref{fig:contour} are well separated from each other, $J_{L/2}$
takes a constant value, which we consider a hallmark feature of ballistic dynamics \cite{langer11}.

However, in one dimension, there is a subtlety as certain solutions of the diffusion equation can also give rise to 
a linear increase of the radius with time (if properly defined).
 Such a scenario happens in the  dilute limit (which we do not study here), yet it results in a strong dependence of the expansion velocity on the total particle, which 
is clearly different from our case as we shall see below. Further details are given  in Appendix A.

The observation of a linear increase of the cloud radius with time implies that
$V_{\mathrm{ex}}$ should be fully determined by properties of the initial state, such as the MDF, energy per particle, or density.
In the non-interacting case, this is obvious, since $V_{\mathrm{ex}}$ can be calculated from the knowledge of the MDF. 
To guide the  interpretation of the interacting case and to understand the dependence of $V_{\mathrm{ex}}$ on $U$ and $n_{\mathrm{init}}$, we next study the two exactly solvable limits $U=0$ and $U=\infty$.

\subsection{Box trap, $V_{\mathrm{ex}}$ at $U=0$}

At $U=0$, opening the trap simply means that particles will 
propagate with a  velocity $v_k=2J\sin(k)$ with a probability given by  
 the MDF $n_k$ in the initial state, which is
$n_k = (1/N) \sum_{l,m,\sigma} e^{-i(l-m)k} \langle c^{\dagger}_{l\sigma} c_{m\sigma}\rangle\,$.
The momenta are chosen to match the open boundary conditions in the box, {\it i.e.}, $k=\frac{\pi l}{L_{\mathrm{init}}+1}; l=1,\dots,L_{\mathrm{init}} $.
By a straightforward evaluation of $R_n^2(t)$ from Eq.~\eqref{eq:rn} and using
the time-dependence of creation and annihilation operators, known exactly at $U=0$,
we obtain $V_{\mathrm{ex}}$ as the average velocity of all particles in the initial state:
\begin{equation}
V_{\mathrm{ex}}^2 = \frac{1}{N} \sum_{k} v_k^2 \, n_k\,.\label{eq:vsc}
\end{equation}
In the $U=0$ case, the initial MDF thus completely determines
the expansion velocity. However, this is an over-complete set of constraints: For a very large $N$,
where boundary conditions cease to matter, we can evaluate
Eq.~\eqref{eq:vsc} analytically:
\begin{equation}
V^2_{\mathrm{ex}} =2J^2 [k_F - \cos(k_F) \sin(k_F)]/k_F\,,
\label{eq:vsc2}
\end{equation}
which yields the full dependence on the initial density at $U=0$ through $k_F\propto n_{\mathrm{init}}$ alone. We can interpret Eq.~\eqref{eq:vsc2} in two ways: If $U=0$, $k_F=\pi n_{\mathrm{init}}/2$, whereas for $U=\infty$, $k_F=\pi n_{\mathrm{init}}$.
Using ED, we have verified the validity of Eq.~\eqref{eq:vsc2} by extracting $V_{\mathrm{ex}}$
from the time-dependence of $R_n(t)$ for $N\sim 160$ (see Fig.~\ref{fig:FSED} in Appendix B).

\begin{figure}[t]
\includegraphics[width=0.45\textwidth]{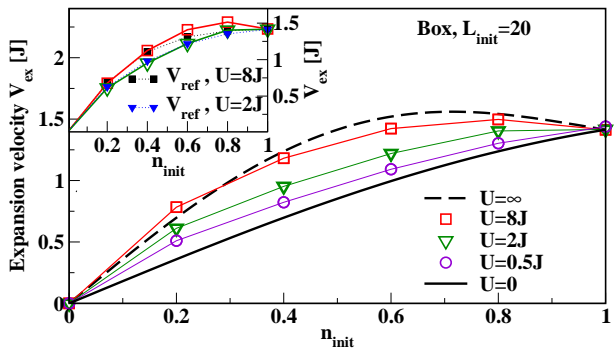}
\caption{(Color online)
Box trap: Main panel: $V_{\mathrm{ex}}$ vs
  $n_{\mathrm{init}}$ at $U/J=0,0.5,2,8,\infty$ for $L_{\mathrm{init}}=20$ [see the legend, symbols are tDMRG, the solid and the dashed lines are derived from Eq.~\eqref{eq:vsc2}].
Inset: $V_{\mathrm{ex}}$
from tDMRG (open squares: $U=8J$, open triangles: $U=2J$) vs. $V_{\mathrm{ref}}$ (solid symbols) from non-interacting reference systems at a finite temperature (see Sec.~\ref{sec:3c} for details) }
\label{fig:vex}
\end{figure}

\subsection{Box trap, $V_{\mathrm{ex}}$ at $U\not=0$} 
In the interacting case, we extract the expansion velocity $V_{\mathrm{ex}}$ from the tDMRG data ({\it i.e.}, the slope of curves such as the ones shown in Fig.~\ref{fig:radius}).
The results for selected values of $U$ are collected in the main panel of Fig.~\ref{fig:vex} (symbols). We emphasize
four  main observations: (i) For  the expansion from the MI, we obtain $V_{\mathrm{ex}}=\sqrt{2}J$
at any  $U>0$. (ii) At a fixed density, $V_{\mathrm{ex}}$ increases monotonically with $U$. (iii) For $U>4J$, the maximum
of the expansion velocity  is at an incommensurate density $0.5<n_{\mathrm{init}}<1$.  
(iv) The expansion velocity is always very different from
 characteristic velocities of the initial state and much smaller than $2J$, the largest possible velocity. It is also much smaller than the charge velocity \cite{giamarchi}  
at small densities and at $n_{\mathrm{init}}=1$,
where the charge velocity drops to zero, $V_{\mathrm{ex}}$ remains finite.

At $U=0$, the first observation is a consequence of particle-hole symmetry, reflected in the MDF: 
$n_k$ is point-symmetric about the point $(k_F= \pi/2,n_{k_F})$. Since $v^2_{k_F+\delta k_F}=v^2_{k_F -\delta k_F}$, from Eq.~\eqref{eq:vsc},
we conclude $V_{\mathrm{ex}} =\sqrt{2}J$. The MDF at $U>0$ has the same symmetry property, hence we expect a similar
behavior, confirmed by tDMRG. Of course, Eq.~\eqref{eq:vsc} does not directly apply to the interacting case.
Since the total energy $E_U=\langle H_0\rangle $ is conserved, for $U>0$, Eq.~\eqref{eq:vsc} is incompatible with this initial condition set by $U>0$ and $n_{\mathrm{init}}$.
However, we shall see that the observation of $V_{\mathrm{ex}} =\sqrt{2}J$ for $U>0$ can also be understood as a consequence of symmetry properties.

We can further use the exact result Eq.~\eqref{eq:vsc2} to  explain the observations (ii)-(iv). The $U=0$ and the $U=\infty$ result are the solid and the dashed lines in the main panel of Fig.~\ref{fig:vex}, respectively, and therefore,  increasing $U$ from $U=0$ 
to $U=\infty$ at a fixed density simply takes us from the limit of a non-interacting {\it two-component} Fermi gas to the limit of non-interacting {\it spinless} fermions.
To understand  that the maximum of $V_{\mathrm{ex}}$ is at an incommensurate $n_{\mathrm{init}}$ for $U>4J$,  one needs to take into account that on the one hand, in a 1D cosine band, the maximum velocity  
is at $k=\pi/2$, but on the other hand, the density of states takes its minimum there. As a consequence of this competition, {\it i.e.}, the decrease of $v_k$ vs the 
increase of the density of states as one moves away from $\pi/2$, the largest expansion velocity is at $n_{\mathrm{init}}\not= 1$. 
Finally, property (iv) is a consequence of {\it all} particles propagating and not just those with momenta close to $k_F$.

On a technical note, we have checked the dependence of $V_{\mathrm{ex}}$ on particle number, keeping $n_{\mathrm{init}}=N/L_{\mathrm{init}}$ fixed.
Finite-size effects are the largest at small initial densities, yet for densities $n_{\mathrm{init}}\gtrsim 0.5$, our tDMRG
results obtained with $L_{\mathrm{init}}=40$ show little quantitative differences compared to smaller $L_{\mathrm{init}}$ and $V_{\mathrm{ex}}$ becomes {\it independent} of $N$ as shown in Appendix B.

\subsection{Reference systems} 
\label{sec:3c}
It is now a compelling question to ask how many constraints suffice to determine the expansion velocity.
From the solution of the non-interacting case, we conclude that density and energy are  relevant quantities.
To check this conjecture for the interacting case, we construct non-interacting reference systems that are at a {\it finite}
temperature \cite{rosch-comm}. The temperature is chosen such that the reference system has the same
energy as the interacting system and the same particle number, and it lives in the same box potential of length $L_{\mathrm{init}}$. 

\begin{figure}[tb]
\includegraphics[width=0.45\textwidth]{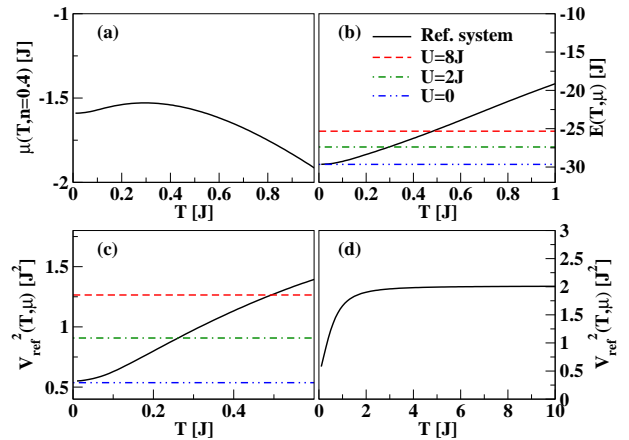}
\caption{(Color online)
These plots illustrate how the non-interacting reference systems are constructed using Eqs. (5), (6) and (7) for the example of  $n_{\mathrm{init}}=0.4$ for (a),(b),(c): $L_{\mathrm{init}}=40$ and (d): $L_{\mathrm{init}}=160$: (a) Temperature dependence of the chemical potential at fixed $n_{\mathrm{init}}=0.4$. (b) Total energy from Eq.~(6) as a function of temperature. (c) $V_{\mathrm{ref}}^2$ as a function of temperature. The horizontal lines in (b) are DMRG results for $E_U$  for the initial states used in the expansion at zero temperature for $U=0,2J,8J$, while in (c) they mark the resulting $V^2_{\mathrm{ref}}=V^2_{\mathrm{ref}}(E,n)$. (d) As $T$ increases, $V_{\mathrm{ref}}^2\to 2J^2$ from below, {\it i.e.}, this is the largest expansion velocity that the reference systems can produce for $n_{\mathrm{init}}\leq 1$.}
\label{fig:Tref}
\end{figure}

Hence we solve
this set of equations:
\begin{eqnarray}
N    &= & \sum_{k,\sigma} f(\epsilon_k-\mu, T) \label{eq:ref1}\,, \\
E_U &= & \sum_{k,\sigma} \epsilon_k f(\epsilon_k-\mu,T)\,, \label{eq:ref2} \\
V_{\mathrm{ref}}^2   &=& \frac{1}{N} \sum_{k,\sigma} v_k^2 f(\epsilon_k-\mu,T)\,,\label{eq:ref3}
\end{eqnarray}  
where $f(x,T)$ is the Fermi function. We proceed as illustrated in Fig.~\ref{fig:Tref}: For a given $U$ and $N$, 
we compute the total energy $E_U$ in the initial state with DMRG. First, we find the chemical potential $\mu=\mu(T)$ from Eq.~\eqref{eq:ref1}, which only depends on $N$.
Using this $\mu(T)$ curve, we  determine the pair of $(\mu,T)$, for which we get the right energy $E_U$.
From these results, Eq.~\eqref{eq:ref3} yields the expansion velocity $V_{\mathrm{ref}}$ of the reference system. 
Obviously,
the maximum velocity that these reference systems, which have the dispersion $\epsilon_k=-2J\cos(k)$ of the empty lattice, can produce is $V_{\mathrm{ref}}=\sqrt{2}J$ at any density $n_{\mathrm{init}}\leq 1$
as $T\to \infty$. Within that constraint, the agreement between $V_{\mathrm{ex}}$ and our reference systems
is excellent, as we illustrate for $U/J=2$ and $8$ in the inset of Fig.~\ref{fig:vex}: Apart from those densities
for which, at $U=8J$, $V_{\mathrm{ex}}>\sqrt{2}J$, $V_{\mathrm{ref}}\approx V_{\mathrm{ex}}$ within our numerical accuracy. 
In the particular case of $n_{\mathrm{init}}=1$, our reference systems also yield $V_{\mathrm{ref}}=\sqrt{2}J$ independently of $E_U$, consistent
with the tDMRG results of Fig.~\ref{fig:vex}. This is a consequence of the aforementioned symmetry property of the
MDF, which also applies to $T>0$.

\section{Expansion from a harmonic trap}
\label{sec:trap}
Our results so far establish a relation between properties of the initial state and  the expansion velocity that could be probed in 
experiments. We next test the robustness of our predictions for $V_{\mathrm{ex}}=V_{\mathrm{ex}}(U,n_{\mathrm{init}})$ against the inhomogeneity induced by a harmonic
potential.

We focus on three types of initial states: (i) Only a TL, {\it i.e.}, $\langle n_i \rangle<1$ in the entire trap, (ii) a MI shell
in the center, surrounded by TL wings, and (iii) a three-shell structure with an incommensurate density in the center $\langle n_i\rangle >1$, 
surrounded by first, a MI shell and second, a TL shell with $\langle n_i \rangle<1$. For a given $U>0$, these regimes are separated by critical characteristic densities $\rho_{1}$ and $\rho_{2}$,
where $\rho=N\sqrt{V/J}$ is the effective density in a system with a harmonic trap \cite{bloch08,rigol04b}. 

\begin{figure}[tb]
\includegraphics[width=0.40\textwidth]{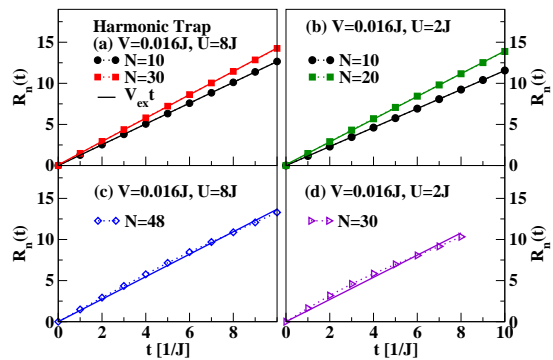}
\caption{(Color online) Harmonic trap: Radius of the particle cloud for the expansion from a harmonic trap for (a) $N=10,30$ at $U=8J$, (b) $N=10,20$ at $U=2J$, (c) $N=48$ at $U=8J$ and (d) $N=30$ at $U=2J$. For $U=8J$, these parameters correspond to the initial states shown in the insets of Fig.~\ref{fig:trap}. The solid lines are fits to the data, dotted lines are guides to the eye. We find $R_n(t)\not\sim t$ whenever densities in the center of the trap 
are larger than one and $R_n(t)\sim t$ otherwise.
}
\label{fig:trapradius}
\end{figure}

\begin{figure}[t]
\includegraphics[width=0.45\textwidth]{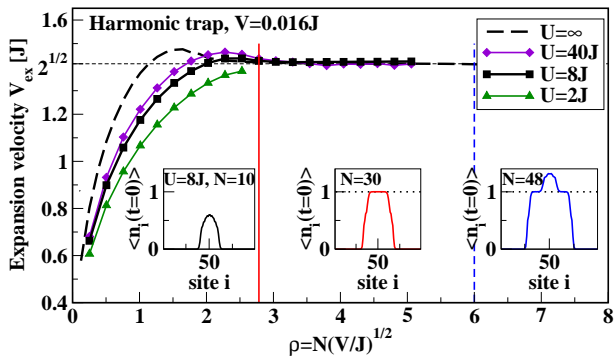}
\caption{(Color online)
Harmonic trap:  $V_{\mathrm{ex}}$ vs $\rho$ ($U/J=2,8,40,\infty$, $V=0.016J$). The vertical solid line marks the 
 formation of a MI shell in the
trapped system at $\rho_1$ and the vertical dashed line the formation of a core with $\langle n_i\rangle >1$ at $\rho_2$, both for $U=8J$. 
The horizontal line is $V_{\mathrm{ex}}=\sqrt{2}J$. 
Symbols are tDMRG results, the dashed line was obtained from ED. 
We have  verified 
that the results are remarkably stable against changes in the
particle number at fixed $\rho$ by producing the $V_{\mathrm{ex}}=V_{\mathrm{ex}}(\rho)$ curve at a different $V$ (see Fig.~10 in  Appendix B).
Lower insets:  typical initial density profiles in the regimes $\rho<\rho_{1}$, $\rho_1<\rho<\rho_2$ and $\rho_{2}<\rho$ for $U=8J$.}
\label{fig:trap}
\end{figure}

For all three cases, $R_n(t)$  is shown in Fig.~\ref{fig:trapradius} for $U/J=2$ and $8$.
We observe that, after releasing the particles from the harmonic trap, the cloud still expands with $R_n(t) \sim t$ in cases (i) and (ii), {\it i.e.},
$R_n(t)\sim t$ [see Fig.~\ref{fig:trapradius} (a) and (b)] whereas in case (iii), the increase of the radius is slower than linear in $t$ [see Fig.~\ref{fig:trapradius} (c) and (d)]. In that  regime and for $U>4J$,
the system can be viewed as a mixture of single atoms propagating with velocities $v_k \sim J$ and two fermions repulsively bound into a doublon, which, due to energy conservation,
does not decay on time scales $\propto 1/J$ and is much slower with typical velocities $v_k^{d}\sim J^2/U$ \cite{kinoshita06,hm09a}. 
For illustration, the values of  $\rho_{1}$ and $\rho_{2}$ as well as typical density  profiles are included in  Fig.~\ref{fig:trap} for $U=8J$ (vertical lines 
and lower insets, respectively).
As is evident from Fig.~\ref{fig:trap}, the overall dependence of $V_{\mathrm{ex}}=V_{\mathrm{ex}}(\rho)$ resembles that of the 
expansion from a box trap, with a maximum  in $V_{\mathrm{ex}}$ emerging as $U\gtrsim 4J$. 
Most importantly, as soon as the MI  forms in the center of the trap, indicated by the
vertical solid line at $\rho=\rho_1$, the expansion velocity approaches  a constant value  at $V_{\mathrm{ex}}\gtrsim \sqrt{2}J$ from above. 
The contribution  to $V_{\mathrm{ex}}$ of low-density shells surrounding the MI is suppressed by increasing $U$ or $\rho$ since both 
favor a  large relative fraction of all particles in the MI shell to minimize the contribution from the 
interaction energy. 
In contrast to the 
expansion from the box, the limit of $U=\infty$ (dashed line) is approached very slowly since the shell structure in a trap 
depends strongly on $U$ and $\rho$.

\section{Summary}
\label{sec:summary}
We studied the sudden expansion of 
a spin-balanced two-component gas in 1D,  released from a trap. Our main results are two-fold: 
First, the cloud expands ballistically as long as initial densities are small,
including, in particular, the MI state. 
Second, the expansion velocity, defined through 
$R_n(t)=V_{\mathrm{ex}}t$ strongly depends on initial density and thus, its measurement can provide information on the initial state.
For instance, deviations from our predictions could indicate the presence of defects in the initial state preparations. 
Our quantitative predictions can be tested in an experiment that 
realizes the set-up of Ref.~\cite{schneider11} in 1D.

Furthermore it would be interesting to study the radius of an expanding cloud and the expansion velocity for other experimentally relevant systems such as the Bose-Hubbard model or spin imbalanced mixtures.
While we have presented phenomenological evidence for ballistic dynamics, we have here not touched upon a potential
relation with integrability and non-trivial conservation laws \cite{zotos97}, leaving this for  future research.
It also remains as an open question to identify interacting models in one dimension and parameter regimes 
in which diffusive dynamics dominates during  the sudden expansion, which might be challenging since even non-integrable
models may have very large conductivities (see, {\it e.g.}, Ref.~\cite{rosch00}).

\begin{acknowledgments}
We thank A. Feiguin, M. Rigol, A. Rosch, and U. Schneider  for very helpful discussions.
F.H.-M. and U.S. thank the KITP at UCSB, where this work was initiated, for its hospitality. This research
was supported in part by the National Science Foundation under Grant No. NSF PHY05-51164.
S.L., M.S., F.H.-M., and U.S. acknowledge support from the DFG through FOR 801. Ian McCulloch acknowledges support from the Australian Research Council Centre of Excellence for Engineered Quantum Systems.
\end{acknowledgments}

\section*{Appendix A: Linear increase of the radius from a nonlinear diffusion equation}
Here we discuss solutions of the diffusion equation in one dimension in the limit of a very dilute gas. 
Since the sudden expansion scenario considered in this paper involves the propagation of all particles, 
the dependence of the diffusion constant $D$ on the local density $n(x,t)$ becomes relevant, and as a consequence, the relevant diffusion equation is in general a nonlinear one (see, e.g., \cite{schneider11}). 
Focussing on the very dilute limit we use $D\sim1/n(x,t)$ (see the discussion in Ref.~\cite{schneider11,footnotediff}). The resulting diffusion equation (after rescaling of the time variable ):
\begin{equation}
\partial_t n(x,t)=\partial_x\frac{1}{n}\partial_x n\,, \label{eq:diffeq}
\end{equation}  
 has a self-similar solution with particle number conservation in 1D \cite{Vazquez06}:
\begin{equation}
n(x,t)=\frac{2t}{x^2+v^2t^2}\,.
\label{eq:ndiff}
\end{equation}

\begin{figure}[tb]
\includegraphics[width=0.45\textwidth]{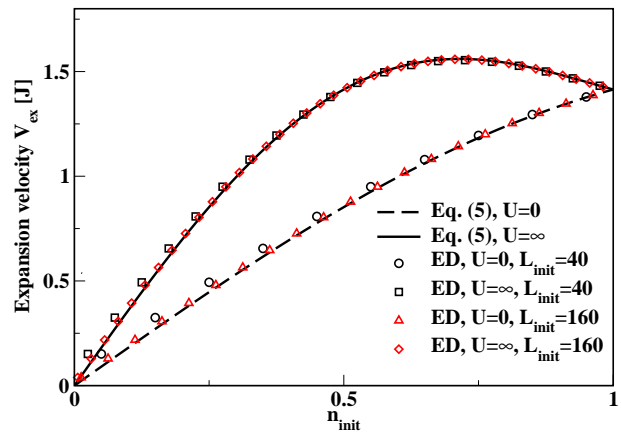}
\caption{(Color online) Box trap: Comparison between the exact result for $V_{\mathrm{ex}}$ [Eq.~(4) of the main text, valid in the limit of large particle numbers] and exact diagonalization in the limits $U=0$ and $U=\infty$. ED data are shown for $L_{\mathrm{init}}=40$ and $L_{\mathrm{init}}=160$.  In the latter case, we find very good agreement between the analytical result (lines) and the ED data (symbols) for all densities.}
\label{fig:FSED}
\end{figure}

\begin{figure}[tb]
\includegraphics[width=0.45\textwidth]{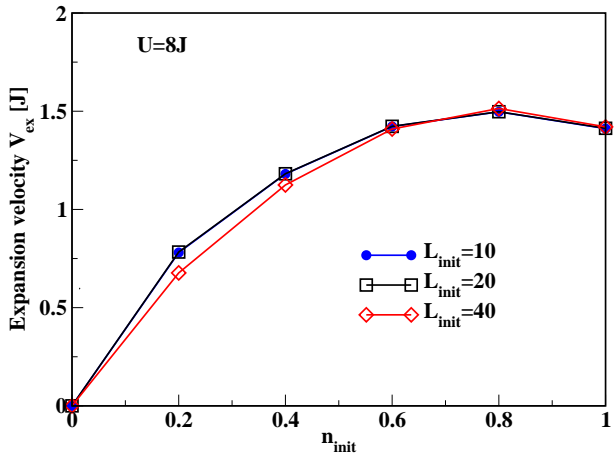}
\caption{(Color online) Box trap: Expansion velocity as a function of initial density for $U=8J$ and different $L_{\mathrm{init}}=10,20,40$. For $n_{\mathrm{init}}\gtrsim0.6$, finite-size effects are remarkably small.}
\label{fig:FSbox}
\end{figure}

First of all, one realizes that our definition of the radius $R_n(t)$, Eq.~\eqref{eq:rn}, cannot be used here. In the analysis of experimental data, it is common practice 
to define the radius as the half-width at half-maximum of the expanding cloud \cite{schneider11}. 
Using this definition, the solution Eq.~\eqref{eq:ndiff}  yields indeed $R_n(t)=vt$, similar to the ballistic dynamics discussed in our work.  
We would like to stress, though, that the sudden expansion described in the main text is genuinely different in some important respects.  
First, Eq.~\eqref{eq:diffeq} is only valid in the dilute limit while the time-dependent DMRG gives us access to short and intermediate time-scales only where the gas is not necessarily a dilute one yet. 
Second, Eq.~\eqref{eq:ndiff} is  a solution for which the expansion velocity $v$  depends strongly on the particle number via $N=\int_{-\infty}^{\infty} n(x,t) dx=2\pi/v$, 
which is not observed in our case (compare Fig.~\ref{fig:vex}~and~\ref{fig:FSbox}). 
Based on these differences, we conclude that diffusive dynamics is very unlikely to be realized for the 1D Hubbard model in the sudden expansion.

\begin{figure}[t]
\includegraphics[width=0.45\textwidth]{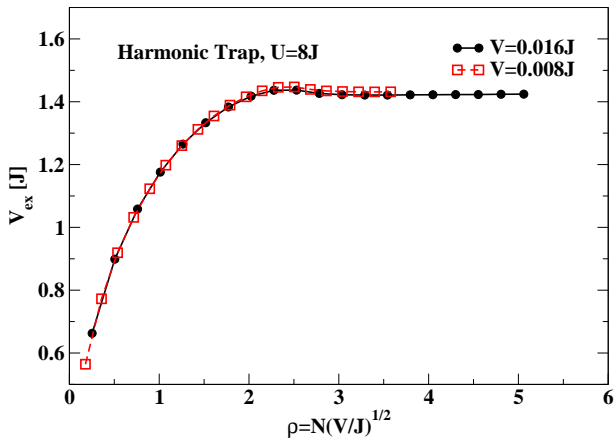}
\caption{(Color online) Harmonic trap: Expansion velocity for $V=0.008J$ and $V=0.016J$ as a function of effective density $\rho=N\cdot\sqrt{V/J}$ for $U=8J$. The expansion velocity is remarkably stable against changing the particle number at fixed $\rho$.}
\label{fig:FStrap}
\end{figure}

\section*{Appendix B: Finite-size effects}

Here we address the question of how our results for the expansion velocity depend on the overall particle number at a fixed density $n_{\mathrm{init}}$. First, we consider the box trap and we compare our analytical result for large $N$ [Eq.~\eqref{eq:vsc}] to exact diagonalization in the noninteracting limits in Fig.~\ref{fig:FSED}. For $N=40$, we find good qualitative agreement with small finite-size effects, which are the most pronounced for $n_{\mathrm{init}}<0.5$. For $N=160$, the deviations between the analytical expression for $N\to\infty$ and data for a finite $N$ are already barely visible except for very low densities. Second, we study the interacting system expanding from different box traps with $L_{\mathrm{init}}=10,20,40$ at a fixed density for $U=8J$. Figure~\ref{fig:FSbox} shows $V_{\mathrm{ex}}$ as a function of density. As in the non-interacting case finite-size effects are remarkably small whenever $n_{\mathrm{init}}\geq0.6$ even for the smaller particle numbers.
Finally, we turn to the expansion from a harmonic trap and analyze $V_{\mathrm{ex}}$ for two different trapping potentials, $V=0.008J$ and $V=0.016J$. Fig.~\ref{fig:FStrap} shows $V_{\mathrm{ex}}$ as a function of effective density $\rho=N\sqrt{V/J}$. We find that the expansion velocity is very robust against changing the particle number at a fixed $\rho$. Overall, our results for the expansion velocity exhibit only minor finite-size effects in all studied cases.

\end{document}